\documentclass[%
 reprint,
superscriptaddress,
 amsmath,amssymb,
 aps,
prb,
]{revtex4-1}

\usepackage{graphicx}
\usepackage{dcolumn}
\usepackage{bm}
\usepackage{amsmath,amssymb}
\usepackage{graphicx}
\usepackage{color}

\newcommand{\blue}{\textcolor{blue}}

\usepackage{hyperref}

\begin{document}

\title{Au-decorated black TiO$_2$ produced via laser ablation in liquid}

\author{S.O.~Gurbatov}
\affiliation{Far Eastern Federal University, Vladivostok, Russia}
\affiliation{Institute of Automation and Control Processes, Far Eastern Branch, Russian Academy of Science, Vladivostok 690041, Russia}

\author{E.~Modin}
\affiliation{CIC nanoGUNE, Donostia, San Sebastian 20018, Spain}

\author{V.~Puzikov}
\affiliation{Far Eastern Federal University, Vladivostok, Russia}

\author{P.~Tonkaev}
\affiliation{ITMO University, St. Peterburg 197101, Russia}

\author{D.~Storozhenko}
\affiliation{Institute of Automation and Control Processes, Far Eastern Branch, Russian Academy of Science, Vladivostok 690041, Russia}

\author{S.~Sergeev}
\affiliation{Institute of Automation and Control Processes, Far Eastern Branch, Russian Academy of Science, Vladivostok 690041, Russia}

\author{N.~Mintcheva}
\affiliation{Department of Chemistry, University of Mining and Geology, 1700 Sofia, Bulgaria}

\author{S.~Yamaguchi}
\affiliation{Department of Physics, Tokai University, Hiratsuka, Kanagawa 259-1292, Japan}

\author{N.~Tarasenka}
\affiliation{B. I. Stepanov Institute of Physics, Minsk, Belarus}

\author{A.~Chuvilin}
\affiliation{CIC nanoGUNE, Donostia, San Sebastian 20018, Spain}

\author{S.~Makarov}
\affiliation{ITMO University, St. Peterburg 197101, Russia}

\author{S.~A.~Kulinich}
\affiliation{Far Eastern Federal University, Vladivostok, Russia}
\affiliation{Research Institute of Science and Technology, Tokai University, Hiratsuka, Kanagawa 259-1292, Japan}

\author{A.~A.~Kuchmizhak}
\email{alex.iacp.dvo@mail.ru}
\affiliation{Far Eastern Federal University, Vladivostok, Russia}
\affiliation{Institute of Automation and Control Processes, Far Eastern Branch, Russian Academy of Science, Vladivostok 690041, Russia}

\begin{abstract}
Rational combination of plasmonic and all-dielectric concepts within unique hybrid nanomaterials provides promising route toward devices with ultimate performance and extended modalities. However, spectral matching of plasmonic and Mie-type resonances for such nanostructures can only be achieved for their dissimilar characteristic sizes, thus making the resulting hybrid nanostructure geometry complex for practical realization and large-scale replication. Here, we produced unique amorphous TiO$_2$ nanospheres simultaneously decorated and doped with Au nanoclusters via single-step nanosecond-laser ablation of commercially available TiO$_2$ nanopowders dispersed in aqueous  HAuCl$_4$. The fabricated hybrids demonstrate remarkable light-absorbing properties (averaged value $\approx$ 96\%) in the visible and near-IR spectral range mediated by bandgap reduction of the laser-processed amorphous TiO$_2$, as well as plasmon resonances of the decorating Au nanoclusters, which was confirmed by combining optical spectroscopy, advanced electron energy loss spectroscopy, transmission electron microscopy and electromagnetic modeling. Excellent light-absorbing and plasmonic properties of the produced hybrids were implemented to demonstrate catalytically passive SERS biosensor for identification of analytes at trace concentrations and solar steam generator that permitted to increase water evaporation rate by 2.5 times compared with that of pure water under identical one-sun irradiation conditions.

\end{abstract}

\maketitle

\section{Introduction}
Collective resonant oscillations of conduction electrons (also known as localized surface plasmon resonance (LSPR)) in noble-metal nanoparticles (NPs) provide a common and reliable way to control electromagnetic fields at nanoscale. Such oscillations facilitate resonant absorption of incident energy that can be converted to strongly enhanced electromagnetic fields around the NP surface or dissipate via nonradiative damping to induce localized heating. Both electromagnetic and photothermal effects have found numerous practical applications in photovoltaics, solar energy conversion, biomedicine, sensing, etc \cite{giannini2011plasmonic,atwater2011plasmonics,baffou2020applications}. Alternatively, NPs made of various materials with high refractive index (high-$n$) and low dissipative losses in the visible and near-IR spectral range (for example, Si, Ge and TiO$_2$) emerged recently as alternative platform for nanoscale light management \cite{kuznetsov2016optically}. In particular, such nanomaterials support Mie-type resonances that permit to concentrate electromagnetic energy inside the NP bulk giving rise to efficient photothermal conversion and highly enhanced nonlinear optical effects \cite{alessandri2016enhanced,makarov2017efficient,zograf2017resonant,mitsai2019si,koshelev2020subwavelength}.

Rational combination of plasmonic and all-dielectric concepts to design hybrid nanostructures is expected to provide a promising route toward advanced nanomaterials with optimized optical response and extended operation range \cite{jiang2014metal,zuev2016fabrication,timpu2017enhanced,makarov2017light,milichko2018metal}. Meanwhile, spectral matching of LSPRs and Mie-type resonances of noble-metal and high-$n$ NPs is crucial for optimal performance. Unfortunately, in the visible and near-IR spectral ranges such resonant matching can be achieved for dissimilar characteristic dimensions of both types of NPs (in particular, 100-500 nm for dielectric NPs and less than 50 nm - for plasmonic ones). This makes the NP geometry quite complex for practical realization, even with time- and money-consuming lithography-based techniques.

Laser ablation in liquids (LAL) has emerged as a promising high-performance and green approach for nanomaterial preparation \cite{amendola2009laser,zeng2012nanomaterials,zhang2017laser}. When compared with wet-chemistry methods, LAL represents a simple and environmentally friendly technology that can be carried out under normal environmental conditions without external stimuli. Intense pulsed laser radiation generates extremely high local pressures, temperatures and quenching rates, thus providing experimental conditions for production of nanostructures with different phase composition (including unique meta-stable phases) \cite{vailionis2011evidence}, complex chemical composition and morphology \cite{zhang2017formation,shih2018two,alexander2019electronic,tymoczko2019one}. However, so far only a few studies reported on LAL-generated hybrid nanomaterials where plasmonic and dielectric counterparts were combined within practically relevant design \cite{liu2015fabrication,saraeva2018laser,mintcheva2020room}, yet without rigorous assessment of their nanophotonic properties and practical applications.

In this paper, unique amorphous TiO$_2$ spherical-shaped NPs simultaneously decorated and doped with Au nanoclusters were produced via single-step ablation of commercially available TiO$_2$ nanopowders in presence of HAuCl$_4$ with nanosecond (ns) laser pulses. The fabricated hybrids demonstrated remarkable broadband light-absorbing properties (averaged value $\approx$ 96\%) in the visible and near-IR spectral ranges mediated by bandgap reduction of the laser-processed amorphous TiO$_2$, as well as LSPRs of the decorating Au nanoclusters, which was confirmed by combining optical spectroscopy, advanced EELS/TEM and electromagnetic modeling. When placed on a back reflecting mirror, the produced Au@TiO$_2$ NPs demonstrated excellent SERS performance resulted from the coupled Mie resonance of TiO$_2$ spheres and the LSPRs of their decorating Au nanoclusters. Isolated Au@TiO$_2$ NPs were found to provide background-free chemically non-perturbing optical identification of various analytes at initial concentrations down to 10$^{-8}$ M. Au@TiO$_2$ NPs with their strong broadband optical absorption make them promising for photothermal conversion of the solar energy. As a proof-of-concept, by using a commercial cellulose membrane functionalized with solar-energy absorbing Au@TiO$_2$, we realized a lab-scale water steam generator which permitted to increase evaporation rate by 2.5 times compared with that of pure water.

\section{Materials and Methods}
\subsection{Fabrication of Au@TiO$_2$ nanoparticles.}
Two types of commercially available TiO$_2$ nanopowders (anatase powder from Wako Chemicals, 99.99 \% pure, and P25 from Degussa, 99.5 \% pure) with average particle sizes 125 and 21 nm, respectively, were used as supplied. First, NPs were dispersed in deionized water by means of ultrasonication to achieve 0.001\% solution. Then, the suspension (7.5 ml) was transferred to a quartz cuvette (3 x 3 x 6 cm$^3$) and a certain amount of aqueous tetrachloroauric acid (HAuCl$_4$, concentration of 10$^{-3}$ M) was added. Finally, the suspension was magnetically stirred and irradiated with focused nanosecond laser pulses (Quantel Ultra 50, with 7 ns, 532 nm and 20-Hz as pulse width, wavelength and repetition rate, respectively) for a certain period of time at laser fluence of 20 J/cm$^2$.

\subsection{Characterization.}
\textit{Scanning electron microscopy}. SEM images of Au@TiO$_2$ NPs were obtained at accelerating voltage of 5 kV and beam current of 50 pA (Helios NanoLab 450S, Thermo Fisher, USA). Backscattered electron detector was used to achieve atomic number-sensitive contrast for highlighting Au particles.

To obtain information about the internal structure and composition of the Au@TiO$_2$ system, the cross-sections of NPs produced by focused ion beam (FIB) milling (beam current of 43 pA at an accelerating voltage of 30 kV) were studied via SEM imaging. A platinum layer was locally deposited  atop of Au@TiO$_2$ nanospheres to protect their surface structures during FIB milling \blue{(see Supporting Information)}.


More detailed characterization of composition and inner structure was carried out using transmission electron microscopy (TEM) combined with electron tomography. For TEM studies, particles were ultrasonicated in acetone for 5 min and then 30 $\mu$L droplet was placed on copper grid with lacy carbon.
TEM/scanning TEM (STEM) imaging, electron tomography and STEM-EDX (energy dispersive x-ray spectroscopy, EDAX) chemical mapping experiments were performed at an acceleration voltage of 300 kV (Titan 60-300, Thermo Fisher, USA). The microscope was equipped with a monochromated X-FEG and spectrometer Quantum GIF (Gatan, USA) which allowed performing electron energy loss spectroscopy (EELS) with energy resolution better than 50 meV. Probing of the surface plasmons around  Au NPs was done by means of high-resolution EELS technique combined with STEM image acquisition. In this case, the microscope was tuned at a beam energy of 60 keV in STEM mode. The spectrum images were recorded in the low loss region, including zero-loss peak. At each pixel of HAADF stem image, an EELS spectrum was stored with the length of 2048 pixels and energy dispersion 0.01 eV.

Three-dimensional particle morphology was characterized by the electron tomography technique using high angular annular dark-field STEM (HAADF-STEM) imaging mode. The HAADF-STEM regime provides contrast that is strongly dependent on the atomic number (?Z$^2$) in which Au NPs appear much brighter at HAADF-STEM images. Tomographic tilt series were acquired automatically at angles between ?74$^o$ and +74$^o$ at 2$^o$ tilt step. Images were taken with a FEI Tomography 4.0 software in automatic mode; the dwell time for acquisition was set to 2 ?s for the images of 2048 x 2048 pixels with the pixel size of 1.3 nm. The fiducial-less tilt-series alignment and tomographic reconstructions with simultaneous iterative reconstruction (SIRT) techniques were done using in-house Digital Micrograph (Gatan, USA) scripts \cite{rajabalinia2019coupling}. Reconstructed volumes had a voxel size of 5.2 nm
$^3$. The intensity-based segmentation with a local threshold criterion and manual supervision was used. Depending on the intensity value of pixels, they were assumed as belonging to the Au nanoparticles (bright) or TiO$_2$ sphere (grey). Segmentation of Au and TiO$_2$ phases, subsequent 3-D rendering, and statistical calculations were done by using FEI Avizo 8.1 software.

\textit{Optical and Raman spectroscopy}.
Absorption coefficient A=1-R was measured in 200-1700 nm spectral range with an integrating sphere spectrophotometer (Cary Varian 5000) at a spectral resolution of 1 nm. Halogen and deuterium lamps were used as radiation sources for Vis-NIR and UV range, respectively. The bandgap value E$_g$ was determined by the position of the fundamental absorption edge according to Tauc equation:

 \begin{equation}
        (\hbar\nu F(R))^{1/n}=B(\hbar\nu - E_g)
        \end{equation}

where $\hbar$ is a  Planck's constant, $\nu$ is the oscillation frequency of electromagnetic waves, B is a constant showing slope of the linear fit, while F(R) = (1 - R)$^2$/2R represents Kubelka-Munk's function. The value of the exponent denotes the nature of the interband electronic transitions - for direct allowed transitions is equal to n=1/2. The basic procedure for Tauc analysis is to acquire optical absorbance data that cover a range of energies below and above the bandgap transition. E$_g$ value was determined by extrapolating the linear part of the Tauc curve ($\hbar\nu$)$^{1/n}$ on the $\hbar\nu$ axis.

Raman spectra of pristine TiO$_2$ NPs, as well as produced and post-annealed AuTiO$_2$ hybrids, were acquired with commercial Raman microscope (Alpha WiTec) at 532 nm (2.33 eV) pump. Similar device was used to probe the SERS performance of AuTiO$_2$ NPs. The  Au@TiO$_2$ NPs  were  functionalized  with  several  types  of practically relevant  analytes,  namely  organic  dyes  (Rhodamine 6G  and  Acridine  orange),  medical  drags (Diphendramine  hydrochloride) and histological marker molecules (4',6-diamidino-2-phenylindole  dihydrochloride,  DAPI). The NPs were added into the 10$^{-8}$M analyte solutions and stored there for 2 h. Then, suspensions were drop-cast onto a bulk Ag mirror. SERS signal was obtained only from isolated NPs on the substrate and averaged over 50 similar measurements for each analyte to account for random distribution of Au nanoclusters.

\textit{FDTD modeling.}
Finite-difference time domain calculations (Lumerical Solutions Ltd.) were undertaken to assess plasmonic properties of the Au@TiO$_2$ hybrids. The structures were excited  with linearly polarized broadband source. Exact 3D models of the NPs were reconstructed from the corresponding STEM images produced during the EELS studies. The dispersion functions for Au and TiO$_2$ were obtained from Palik \cite{palik1998handbook}.

\subsection{Modelling of radiation-induced heating of Au@TiO$_2$ particles.}

Theoretical description of nanoparticle heating (induced by either solar or laser radiation) was carried out numerically with a commercial software package (Comsol Multiphysics). A numerical model was first constructed for a plane wave irradiation of a bare and Au-decorated ellipsoidal TiO$_2$ NP (the diameter along the long axis is 21 nm) in water. Distribution of electromagnetic fields in the simulated volume excited by a single laser pulse (532 nm wavelength and 25 mJ pulse energy) was calculated by solving wave equation in the frequency domain. Then, the following heat equation in a time domain during the single-pulse irradiation was calculated:
 \begin{equation}
        {\rho C_p\frac{\delta T}{\delta t} + \nabla \cdot (-\kappa \nabla T)=\vec{J}  \cdot \vec{E}},
        \end{equation}
where the right side of this equation describes the heat source with the current density $J$ and electric field amplitude $E$, whereas the left side corresponds to temperature evolution in time and space with corresponding parameters: C$_p$ is the heat capacity at constant pressure, $\rho$ is the material density, and $\kappa$ is the thermal conductivity. We suppose the refractive indexes are $n_{Au}$=0.56+3.44i, $n_{water}$=1.335, $n_{TiO_{2}}$= 2.54+0.0001i. Thermal conductivities for TiO$_2$, gold and water are taken as $\kappa_{TiO_2}$=7 $W/(m \cdot K)$, $\kappa_{Au}$=320 $W/(m \cdot K)$ and $\kappa_{water}$=0.6~$W/(m \cdot K)$. Similar procedure was used to calculate heating efficiency of the Au@TiO$_2$ hybrids of variable diameter in water and on the glass substrate. For these calculations, an excitation source with central wavelength at 550~nm and intensity 0.1~W/cm$^2$ was considered providing reliable approximation of solar irradiation. We define thermal boundary conditions as a heat flux through the computational domain surface $q = h(T_{ext}-T)$, where $h$ is  the heat transfer coefficient, $T_{ext}$ = 293~K is the external temperature, $T$ is the calculated temperature near boundary of the computational domain. We assume that temperature gradient is weak because in real experiment there are a lot of particle nearby. From these considerations we estimate value of $h$ equals 0.22~W/(m$^2\cdot$K).

\subsection{Water steam generation.}
Nanofluid was prepared by suspending 50 mg of Au@TiO$_2$ in 10 mL of distilled water. Also, similar suspension was filtered through a commercial cellulose membrane (pore diameter of 200 nm) and then dried in vacuum. The produced nanofluid and ``black'' membrane loaded with Au@TiO$_2$ NPs was placed on the surface of distilled water and irradiated by commercial solar simulator at 0.1 W/cm$^2$. A microbalance was used to detect weight loss of water after evaporation. The evaporation rate observed in both cases was compared with that of distilled water. Heating process was visualized with calibrated IR camera. All experiments were performed at 25 $^o$C and relative humidity of 40 \%.

\begin{figure*}[t!]
\centering
\includegraphics[width=1.5\columnwidth]{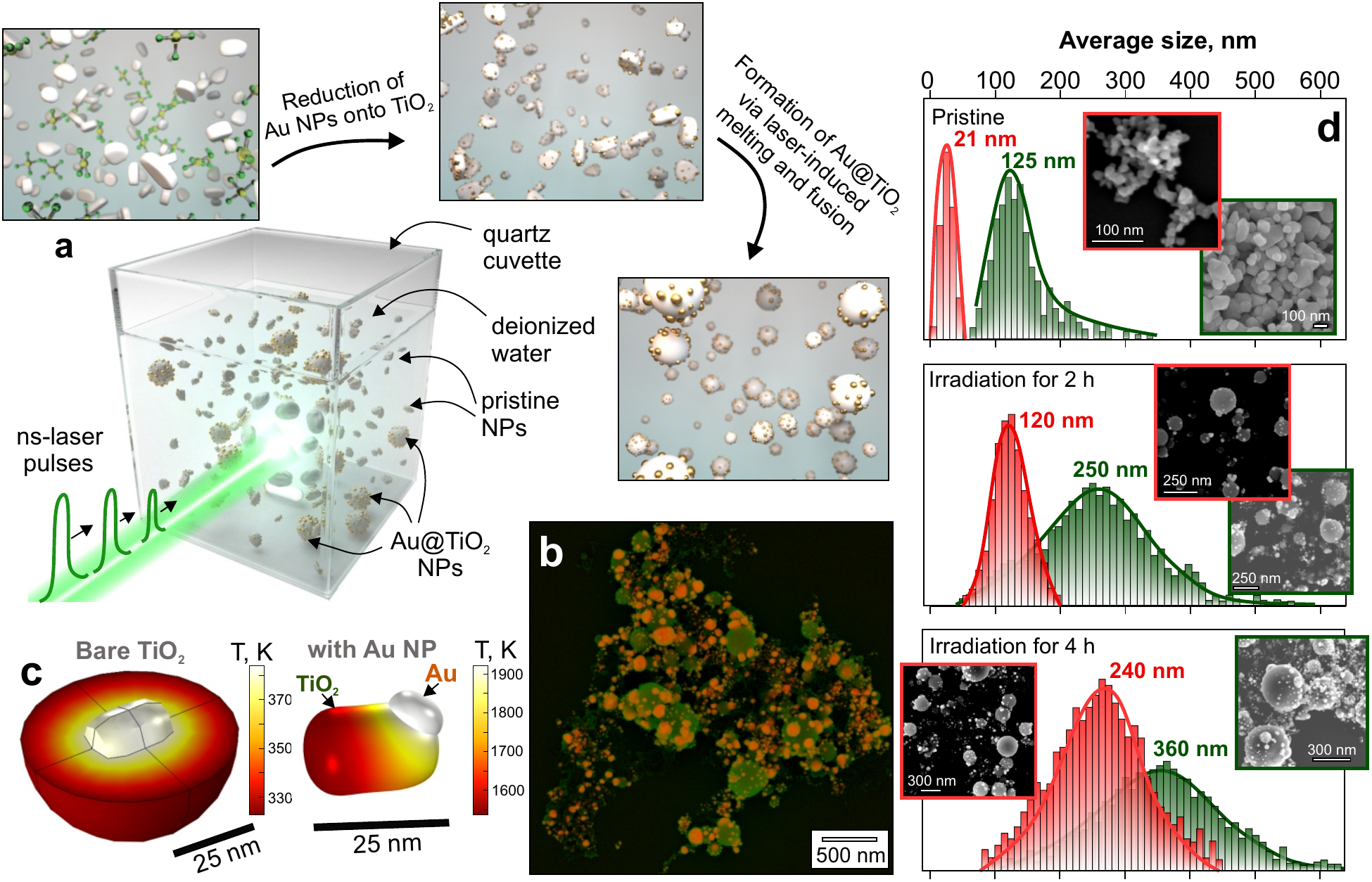}
\caption{\textbf{Fabrication of Au@TiO$_2$ NPs via laser ablation in liquid.} (a) Schematic representation of fabrication process suggesting irradiation of stirred water suspension of TiO$_2$ NPs with ns-laser pulses. Insets explain relevant processes underlying formation of Au@TiO$_2$ NPs. (b) Representative false-color SEM image of NPs produced upon irradiation of suspension with 21-nm-sized TiO$_2$ NPs for 2 h at pulse energy 20 J/cm$^2$ and 20 Hz pulse repetition rate. (c) Calculated temperature profiles in the vicinity of bare and Au-decorated TiO$_2$ NP upon irradiation with single ns-laser pulse at 20 J/cm$^2$. Bare TiO$_2$ NP is shown inside water medium. (d) Size distributions measured for pristine TiO$_2$ NPs of both types (top) and Au@TiO$_2$ product obtained from these NPs after irradiation for 2 and 4 h  (middle and bottom). Each size distribution is illustrated by representative SEM images}.
\label{fig:1}
\end{figure*}

\section{Results and Discussions}
\subsection{Au-decorated black TiO$_2$: fabrication and structural properties.}
Figure 1a schematically illustrates key processes underlying formation of Au@TiO$_2$ NPs using the LAL technique. In our experiments, aqueous suspension of as-supplied TiO$_2$ NPs (mainly anatase, average size of 21 and 125 nm, respectively) mixed with aqueous solution of HAuCl$_4$ was irradiated by ns-laser pulses for a few hours resulting in formation of spherical TiO$_2$ decorated with multiple nano-sized Au clusters (Fig. 1b). Typically, the average size of resulting Au@TiO$_2$ hybrids is always larger than the size of as-supplied NPs (for both types of starting nanopowders). Accordingly, the observed increase in size and spherical shape of produced NPs indicate melting and fusion as key processes involved in their formation.

However, the as-supplied crystalline TiO$_2$ NPs of both types weakly absorb visible light, which results in their single-pulse laser heating that is insufficient to reach the melting point of bulk TiO$_2$ ($\approx$1900 K, \cite{liu2014titanium}). In particular, 21-nm sized TiO$_2$ nanoparticle suspended in water can be homogeneously heated up to only 385 K upon irradiation with a single 7-ns laser pulse at fluence of 20 J/cm$^2$, as confirmed by corresponding numerical modeling (see Fig. 1c and Methods for simulation details). In a sharp contrast, Au nanoclusters can efficiently absorb laser radiation at 532 nm that is close to the wavelength of their localized plasmon resonance. The absorbed radiation is converted to Joule heating by resonantly excited conduction electrons, while the generated heat can be further transferred to surroundings. Similar modeling of the single-pulse laser heating of isolated 21-nm-big TiO$_2$ NP decorated with an Au cluster with its diameter of 5 nm shows that such a hybrid nanostructure can easily reach temperatures exceeding 1900 K at the same irradiation laser fluence (Fig. 1d).

To further support this idea, LAL experiments were performed with a similar suspensions of pristine TiO$_2$ NPs at the same laser fluence and irradiation time but without addition of HAuCl$_4$. Careful SEM analysis of obtained nanomaterials indicated no modification/melting of pristine NPs revealing the key role of Au nanoclusters decorating TiO$_2$ NPs. The formation of such Au NPs can occur even without laser-irradiation and is expected to be preferentially facilitated on TiO$_2$ NPs via surface chemistry driven by active sites and laser-induced enhanced temperature near NP's surface.

\begin{figure*}[t!]
\centering
\includegraphics[width=1.5\columnwidth]{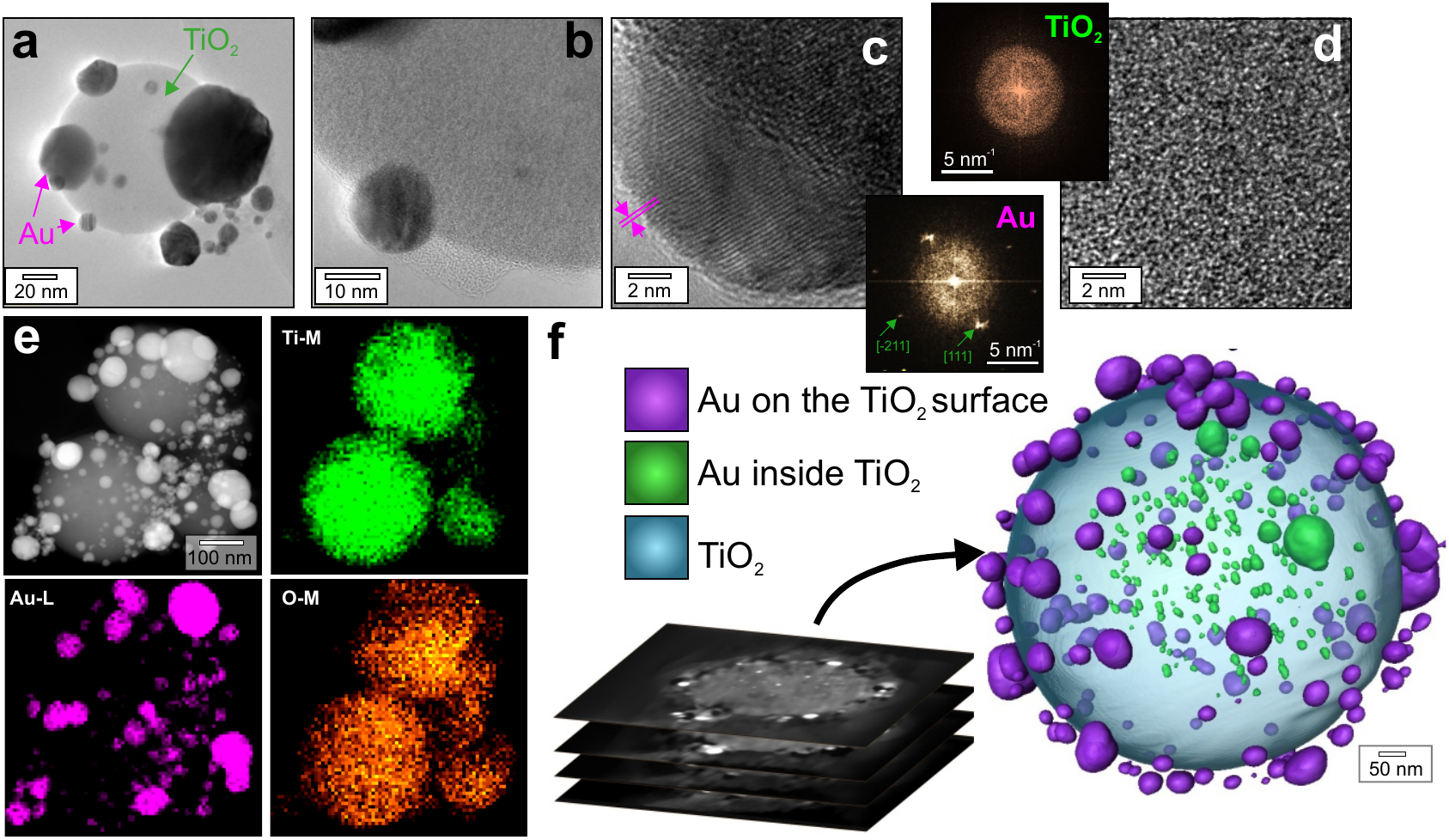}
\caption{\textbf{Structure and composition of laser-fabricated Au@TiO$_2$ NPs}. (a,b) Representative TEM images of Au@TiO$_2$ NPs. (c,d) HR-TEM images showing crystalline structure of an isolated Au nanocluster and TiO$_2$ NP. Insets show the corresponding FFT images. (e) EDX chemical composition mapping of Au@TiO$_2$ NPs. (f) 3D model of an Au@TiO$_2$ NP reconstructed using tomographic series of HAADF-STEM images. For clarity, Au NPs situated on the surface and inside the TiO$_2$ nanosphere are highlighted by purple and green colors, respectively.
}
\label{fig:2}
\end{figure*}

As mentioned above, after irradiation for 2 h all as-supplied TiO$_2$ NPs with irregular shapes were successfully transferred to spherically-shaped Au@TiO$_2$ hybrids. For both types of used pristine NPs with their average sizes of 21 and 125 nm, the obtained hybrids had the averaged diameters of 120 and 220 nm, respectively, indicating that fusion of several NPs is crucial (see Fig. 1d). Shorter irradiation time resulted in particular presence of pristine TiO$_2$ NPs in the obtained product.
However, irradiation of the TiO$_2$ suspension for 4 h yielded in even larger size of obtained Au@TiO$_2$ hybrids. The number of adsorbed Au nanoclusters and their size were also found to increase with laser irradiation time, as well as with the concentration of HAuCl$_4$ in the irradiated dispersion. From practical application point of view, it is important to control both the average size of the spherical-shaped TiO$_2$ NPs and their decoration degree. We showed that by varying the starting size of pristine TiO$_2$ NPs, irradiation time and HAuCl$_4$ content, both the two parameters could be flexibly controlled. The related information is summarized in Fig. 1d and in the \blue{Supporting Information}.

 HR-TEM imaging was used to identify crystalline structure of the obtained Au@TiO$_2$ product. TEM images of representative NPs are shown in Fig. 2(a-d), revealing completely disordered lattice in the TiO$_2$ nanosphere and crystalline structure of decorating Au nanoclusters. FFT analysis of the selected HR-TEM images confirmed amorphous nature of the TiO$_2$ nanospheres (insets in Fig. 3c,d). Additionally, Raman spectroscopy was utilized to provide more information regarding the crystallinity of TiO$_2$ statistically averaged over multiple NPs before and after LAL processing. These studies permitted clearly to identify all main anatase Raman bands for both types of as-supplied TiO$_2$ NPs and low-intense bands at 440 and 610 cm$^{-1}$ that can be attributed to a small amount of nanocrystalline rutile inclusions in the Au@TiO$_2$
 product (\blue{Supporting information}). EDX mapping was used to confirm chemical composition of the Au@TiO$_2$ NPs (Fig. 2e).

 The formation mechanism suggests that Au NPs stimulating TiO$_2$ melting can appear not only on the surface of Au@TiO$_2$ product but also in its bulk upon fusion. To clarify this, several FIB cross-sectional cuts of Au@TiO$_2$ were first visualized using SEM revealing high-contrast nanoscaled inclusions that appeared much brighter compared with TiO$_2$ surrounding and can be attributed to Au \blue{(see Supporting Information)}. More detailed studies of the inner structures of Au@TiO$_2$ NPs were carried out via tomographic reconstruction of HAADF-STEM image series (see Methods for details). The reconstructed 3D model of an isolated Au@TiO$_2$ NP is shown in Fig. 2f, where Au nanoclusters located on the surface and inside the TiO$_2$ NP are highlighted by different colors for clarity. Systematic studies of the produced Au@TiO$_2$ hybrids indicated that the inner Au nanoclusters occupy $\approx$0.5-1\% of the volume of TiO$_2$ nanosphere. Noteworthy, such Au nanoclusters both embedded into TiO$_2$ NPs and simultaneously decorating their surface are reported for the first time.

\subsection{Optical and plasmonic properties of Au@TiO$_2$ hybrids}
We started by probing the plasmonic response of Au nanoclusters capping amorphous spherical-shaped TiO$_2$ using high-resolution EELS. The corresponding EELS spectrum measured from an isolated Au nanocluster sitting on the surface of a 225-nm-sized spherical TiO$_2$ NP demonstrates a pronounced signal at 2.3 eV (Fig. 3a).  Several peaks with their maxima centered at 2.02, 2.145 and 2.365 eV can be resolved indicating the multi-resonant nature of the electron plasma oscillations in the Au nanocluster. Mapping of the intensity of this signal (1.9-2.5 eV) confirmed that it originates from the Au nanocluster surface (Fig. 3b). Noteworthy, the EELS peaks at 5.7 and 10 eV can be attributed to low-loss edges of Ti.

\begin{figure}[t!]
\centering
\includegraphics[width=0.9\columnwidth]{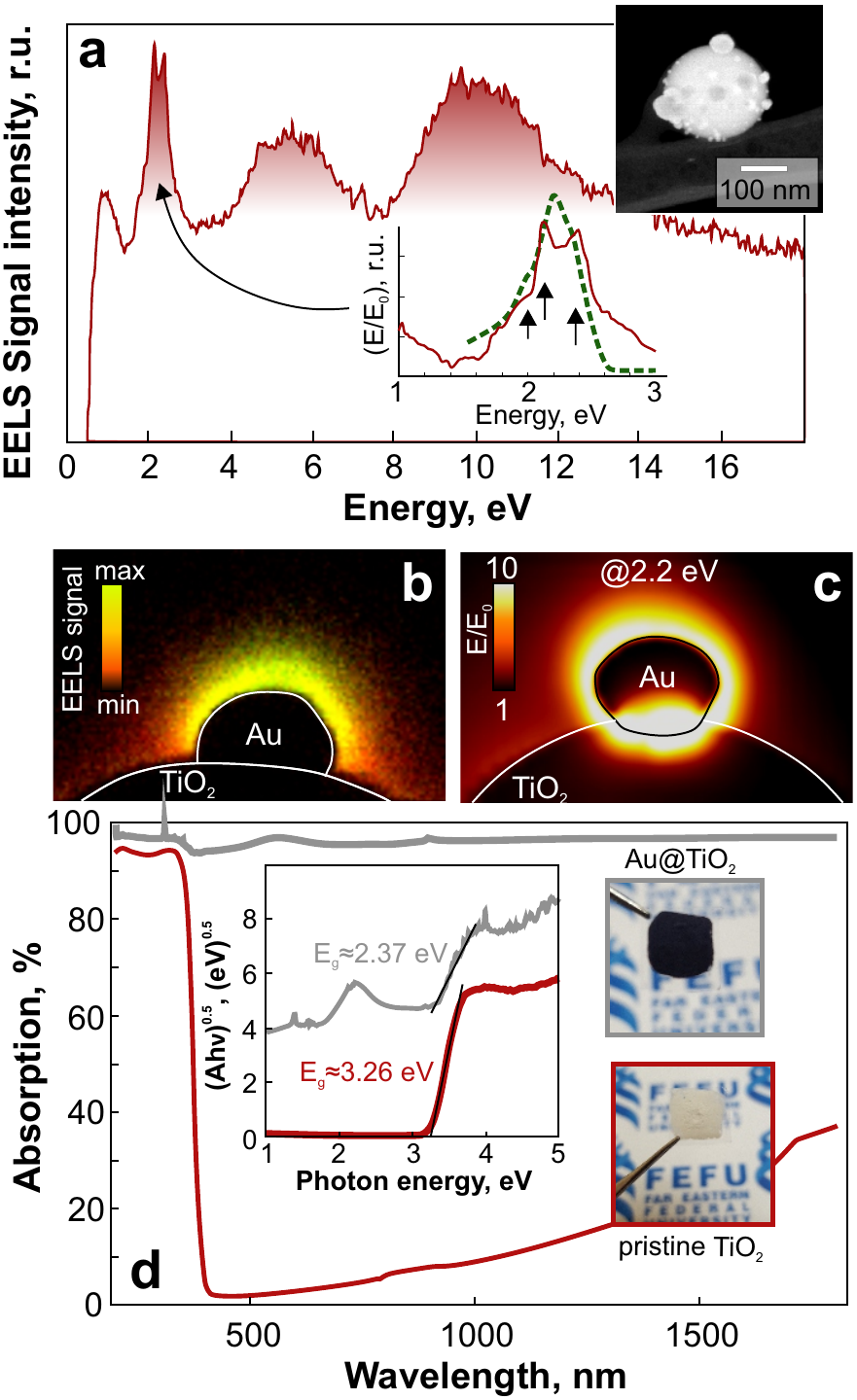}
\caption{\textbf{Optical properties of Au-decorated black TiO$_2$ NPs}. (a) High-resolution EELS spectrum of isolated Au nanocluster capping the TiO$_2$ NP surface. Reference TEM image is shown as upper inset. Bottom inset provides close-up look at EELS spectrum near the plasmon resonance feature superimposed with a calculated averaged near-field plasmonic spectrum of the Au nanocluster (green dashed curve).(b) EELS map showing signal intensity in the spectral range from 1.9 to 2.4 eV. (c) Normalized EM-field amplitude E/E$_0$ calculated at 2.2 eV pump. (d) Absorption spectra of pristine TiO$_2$ and Au@TiO$_2$-based opaque coatings covering the glass slide. The insets show Kubelka-Munk representation of the absorption spectra with an indication of bandgap E$_g$ for Au@TiO$_2$ and pristine TiO$_2$, as well as the optical images of both samples.
}
\label{fig:3}
\end{figure}

The size of the considered Au nanocluster is comparable with a doubled skin depth of gold ($\approx$ 30 nm) thus suggesting the dipolar approximation applicable for formal analysis of its plasmon modes. Considering the dielectric permittivity of Au, the localized dipolar plasmon resonance of such a NP in vacuum should appear around 2.35 eV redshifting to 2.1 eV when it attaches a high-refractive-index TiO$_2$ surface. Also, the shape of the obtained Au nanoclusters (grown above or even penetrating into their TiO$_2$ support; \blue{see Supporting Information}) is typically irregular suggesting a certain splitting of  localized plasmon resonances. The resonance seen at 2.365 eV can be attributed to the quadrupolar localized plasmonic modes appeared owing to the substrate-induced redshift of the dipolar one. These inferences are consistent with the supporting FDTD calculations showing the near-field plasmonic spectrum of the slightly elliptical Au nanocluster capping TiO$_2$ NP as well as the normalized amplitude of the electromagnetic near-fields under resonant excitation (at 2.1 eV; Fig. 3a,c).

Statistical EELS studies averaged over various Au clusters found on TiO$_2$ spheres indicate that the Au@TiO$_2$ hybrids support localized plasmons in a rather broad spectral range spanning from 500 to 650 nm. Moreover, closely spaced Au clusters can act as plasmonic oligomers with their resonance shifted to near-IR part of the spectrum \blue{(see Supporting Information)}.

Further, we assessed optical properties Au@TiO$_2$ hybrids in the visible and near-IR part of their spectra (Fig. 3d). For this purpose, the Au@TiO$_2$ suspension (produced by laser irradiation of P25 pristine TiO$_2$ for 4h) was dried on a cover glass slide to form a uniform opaque coating. In a sharp contrast to the similar coating made of both types of pristine TiO$_2$ NPs, the color of the Au@TiO$_2$ nanomaterial appeared black indicating high absorption in the visible spectral range. Corresponding measurements of absorption coefficient (A=1-R; R - diffuse reflectance) of both coatings revealed an order of magnitude larger vis-to-IR absorption by the ``black'' Au-decorated amorphous TiO$_2$ with respect to its ``white'' pristine precursors. Based on the obtained diffuse reflectance, we found a considerable decrease of the bandgap E$_g$ value from 3.26 (pristine TiO$_2$) to 2.37 eV (Au@TiO$_2$ hybrids) for direct transitions according to Kubelka-Munk method (see inset in Fig. 3d). Noteworthy, the maximal absorption ($\approx$97\%) mediated by the localized plasmon resonances of the Au nanoclusters capping TiO$_2$ was observed at 2.2 eV that is consistent with the numerical modeling and EELS studies.

\subsection{Use of Au@TiO$_2$ hybrids for label-free biosensing and photo-thermal conversion.}
The observed remarkable optical and plasmonic properties of the prepared Au@TiO$_2$ hybrids suggest several areas for their potential application. First of all, the plasmon-mediated EM field localized near Au particles make such Au@TiO$_2$ NPs attractive for label-free biosensing based on the surface-enhanced Raman scattering (SERS) effect \cite{}. As the electromagnetic contribution to the SERS signal is predominant roughly scaling as forth power of the local electromagnetic field amplitude normalized over the pump field amplitude E$^4$/E$_0^4$ \cite{}, maximization of this value gives a general route for boosting SERS performance. Au nanoclusters produce enhanced plasmon-mediated EM fields under resonant visible-light excitation. However, the nanoscale size of such particles results in their low absorption cross-section. As a result, they efficiently interact only with a small fraction of incident pump laser beam with a diffraction-limited lateral size typical for SERS experiments. In the produced Au@TiO$_2$ hybrids, the excitation of decorating Au particles can be improved via additional coupling of the incident field to the Mie resonances supported by TiO$_2$ spheres with diameters comparable with the diffraction-limited ($\approx \lambda$/2) laser spot  \cite{checcucci2018titania, milichko2018metal}. To illustrate this, we performed FDTD calculations showing a 2-fold increase of the local E/E$_0$ amplitude near an isolated Au particle upon its attachment to a 240-nm-sized TiO$_2$ sphere supporting magnetic-type dipolar resonance at plasmon excitation wavelength (2.2 eV). Such improvement permits to expect contribution to SERS electromagnetic enhancement which is at least an order of magnitude larger compared to those from a similar isolated Au particle in a free-space.

\begin{figure}[t!]
\centering
\includegraphics[width=0.85\columnwidth]{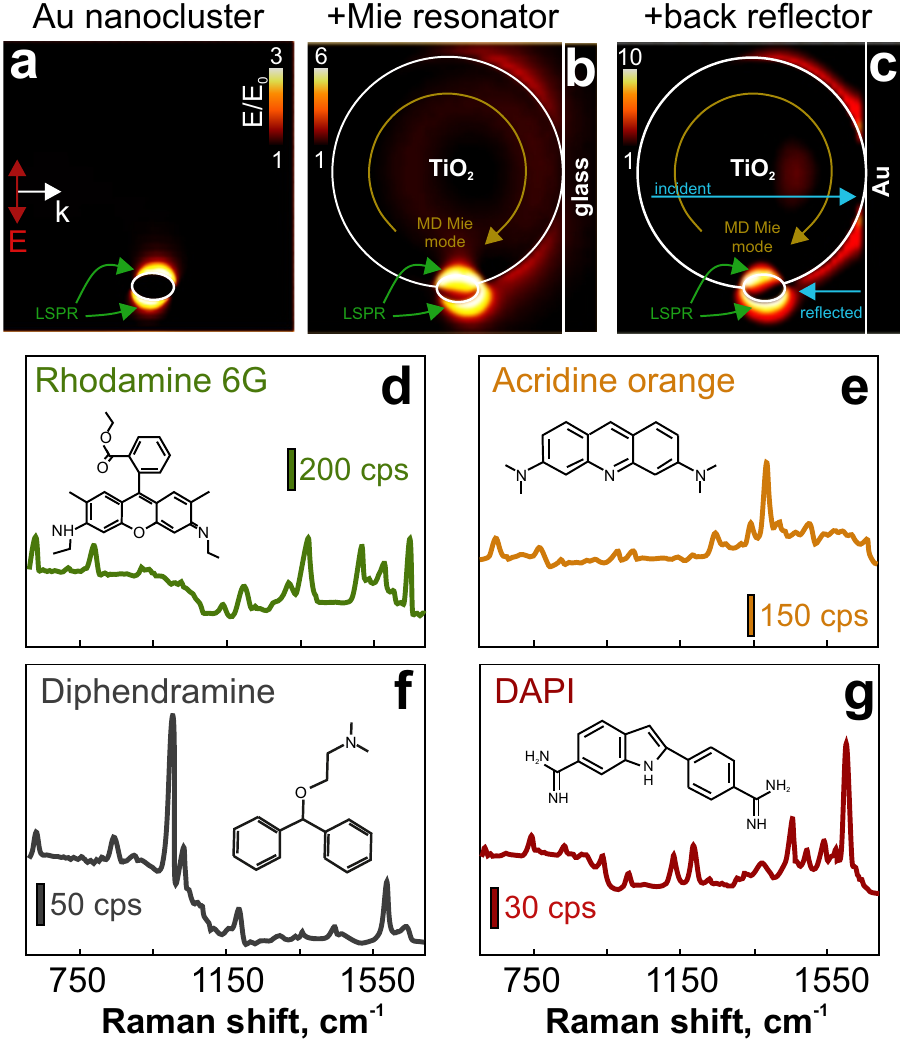}
\caption{\textbf{SERS performance of the Au@TiO$_2$ hybrids on a back-reflecting mirror}. Average enhancement of the electromagnetic field amplitude E/E$_0$ calculated for isolated Au NP in a free space (a), as well as Au NP attached to the 240-nm diameter TiO$_2$ sphere on a glass (b) and Au substrate (c). Series of averaged SERS spectra of Rhodamine 6G (d), Acridine orange (e), Diphendramine (f) and DAPI (g).
}
\label{fig:3}
\end{figure}

Additionally, the geometry of the prepared Au@TiO$_2$ hybrids permits to apply them with back-reflector mirror \cite{}. In our case, this concept can be simply realized upon SERS studies by placing Au@TiO$_2$ NPs on a smooth refectory mirror. In this respect, the TiO$_2$ sphere acts as a support for Au nanoclusters providing required gap between plasmonic NPs and reflecting mirror. Corresponding FDTD calculations of the local EM field enhancement gave more than a 3-fold enhancement of E/E$_0$ (compared with the case of a Au particle in a free-space) near the isolated Au nanocluster yielding in further increase of the SERS performance (Fig. 4c). Amorphous structure of TiO$_2$ spheres contributed to the background free SERS studies.

To assess the applicability of Au@TiO$_2$ hybrids for reliable label-free optical identification at trace concentrations, we probed the SERS signal from isolated NPs with an average diameter 240$\pm$20 nm at 532 nm (2.33 eV) as pumping wavelength. The Au@TiO$_2$ NPs were functionalized with several types of practically relevant analytes, namely organic dyes (Rhodamine 6G and Acridine orange), medical drags (diphendramine hydrochloride) and histological marker molecules (4',6-diamidino-2-phenylindole dihydrochloride, DAPI). Considering the rather random size distribution and arrangement of decorating Au nanoclusters, for each type of the analyte we statistically averaged the SERS signal over at least 50 NPs performing measurements only from isolated structures. By laser pumping isolated Au@TiO$_2$ NPs stored in the analyte solution for 2h and then drop-cast onto a silver mirror, we found distinct SERS signals from all tested analytes (Fig. 4d). The spectral position of the identified Raman bands were found to be in good agreement with previously reported studies substantiating isolated Au@TiO$_2$ hybrids as simple all-in-one SERS platforms with optimized EM response for reliable fingerprinting at initial concentrations down to 10$^{-8}$M \cite{pavliuk2020ultrasensitive}.

Noteworthy, NPs for SERS studies provide additional flexibility for device designs. For example, quantitative SERS measurements using NPs can be performed in liquids to detect and trace catalytic processes \cite{mitsai2018chemically}. Moreover, the analyte solution loaded with functional NPs can be dried on a substrate with specially designed wetting properties. After solvent evaporation on such surface, the deposition area for both the molecules and functional NPs will be substantially reduced increasing local concentration of the analyte already mixed with SERS-active probes \cite{de2011breaking,yang2016ultrasensitive,pavliuk2020ultrasensitive}. Realization of such NP-based sensor active device will become a subject of our forthcoming studies.

Moreover, several studies highlighted the excellent catalytic performance of laser-modified titania again visible-light driven degradation of dye molecules \cite{chen2015laser}. In this work we used methylene blue and methylene orange organic dye molecules to benchmark photocatalytic activity of the produced Au@TiO$_2$ hybrids \blue{(see Supporting information)}. Remarkably, in comparison with the as-supplied pristine TiO$_2$ (P25 from Degussa) the Au@TiO$_2$ hybrids with amorphous core demonstrate weak activity in photocatalytic degradation of both dye molecules. Noteworthy, weak photocatalytic  activity of nanostructres are beneficial in designing SERS substrates for studies where non-invasive and non-perturbing identification of analytes is mandatory \cite{mitsai2018chemically}. It should be noted that post-annealing of the Au@TiO$_2$ product can be used to convert its amorphous core to rutile crystalline phase. Raman spectroscopy indicated appearance of characteristic rutile bands (at 445 and 610 cm$^{-1}$) in Au@TiO$_2$ hybrids after their annealing at 500$^o$ for 2h \blue{(see Supporting Information).} However, more detailed studies of post-annealing of Au@TiO$_2$ NPs and its effect on crystalline structure and photocatalytic activity of the product are out of the scope of this paper.

Finally, the excellent broadband optical absorption of the prepared Au@TiO$_2$ NPs make them promising for photothermal conversion of the solar energy. In particular, the maximal visible light absorption ($\approx$97\%) of Au@TiO$_2$ nanomaterial mediated by the localized plasmon resonances in Au nanoclusters was observed at 2.2~eV which perfectly matches the maximum of the solar energy spectrum. Our numerical calculations showed that Au@TiO$_2$ NPs with diameters ranging from 300 to 500~nm in the form of water suspensions or continuous coatings on a substrate can be efficiently heated under optical excitation that is close to one-sun irradiation conditions (Fig. 5a). Thus, local heating provided by Au@TiO$_2$ NPs allows for fast liquid-vapor phase transition that can be applied for steam generation or water desalination~\cite{zhou20163d,zhao2020materials,neumann2013solar,jin2016steam}.

\begin{figure}[t!]
\centering
\includegraphics[width=1.\columnwidth]{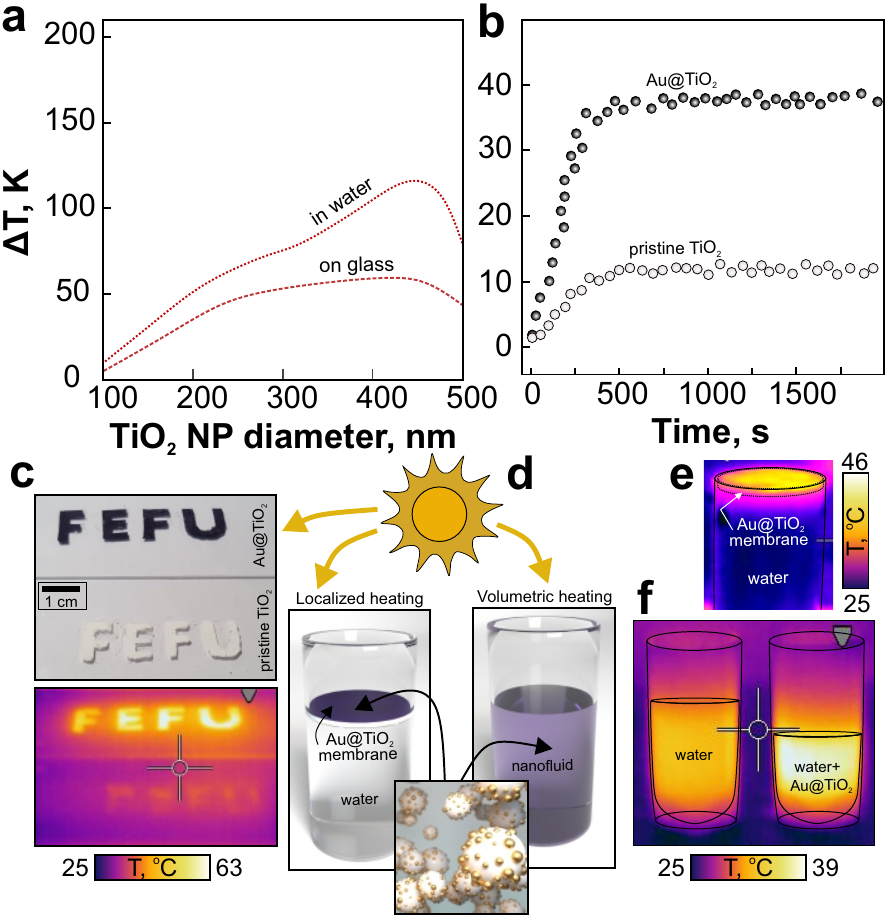}
\caption{\textbf{Photo-thermal conversion with AuTiO$_2$ hybrids}. (a) Calculated temperature increase $\Delta$T in the vicinity of Au@TiO$_2$ NPs of variable diameter D in water and on a glass substrate under conditions close to one-sun irradiation. (b) Measured temperature increase of the coatings made of black Au@TiO$_2$ product and pristine white TiO$_2$ material. (c) Optical and IR thermal images of corresponding coatings arranged to form ``FEFU'' letters. (d) Schematic illustration of volumetric and localized heating with Au@TiO$_2$ hybrids. (e,f) Comparative thermal images of bare distilled water,  Au@TiO$_2$-based nanofluid and distilled water capped with Au@TiO$_2$ membrane.
}
\label{fig:2}
\end{figure}

We performed comparative experiments run with custom-built solar simulator heating the black Au@TiO$_2$ product and pristine white TiO$_2$ material arranged to form ``FEFU'' letters under one-sun irradiation ($\approx$0.1 W/cm$^2$). These experiments showed that the black Au@TiO$_2$ coating reaches a maximal temperature of about 60$^o$C within 5 min in a sharp contrast with white TiO$_2$ reaching only 35$^o$C as indicated by temperature measurements with a calibrated IR-camera (see Fig. 5b).  Further, using the Au@TiO$_2$ product, we realized a lab-scale water steam generator based on both localized and volumetric heating approaches (Fig. 5d). To do this, the weight loss of water during evaporation under one-sun irradiation (at 0.1 W/cm$^2$) was monitored and evaluated for a Au@TiO$_2$-based nanofluid and cellulose-membrane functionalized Au@TiO$_2$ NPs (Fig. 5d). For both approaches, we found that the evaporation rate increases 2-2.5 times compared with that for pristine distilled water, implying the potential applicability of light-absorbing AuTiO$_2$ for solar steam generation and water desalination.

\section{Conclusions}
Here, we produced unique amorphous TiO$_2$ nanospheres simultaneously decorated and doped with Au nanoclusters via single-step ns-laser ablation of commercially available TiO$_2$ nanopowders dispersed in presence of aqueous HAuCl$_4$. The fabricated hybrids demonstrate remarkable light-absorbing properties (averaged value $\approx$ 96\% in the visible and near-IR spectral range mediated by bandgap reduction of the laser-processed amorphous TiO$_2$ as well as LSPRs of the decorating Au nanoclusters that was confirmed by combining optical spectroscopy, advanced EELS/TEM and electromagnetic modeling. Excellent light-absorbing and plasmonic properties of the produced hybrids were implemented to demonstrate catalytically passive SERS biosensing for identification of analytes at trace concentration, as well as solar steam generator that permits to increase water evaporation rate by 2.5 times compared with that of pure water under identical one-sun irradiation conditions. We also envision that the produced Au@TiO$_2$ product will be useful as transport layers in third-generation solar cells, for which our NPs possess excellent optical properties, low cost, and ability to be deposited by scalable wet-chemistry approaches. Similarly, the developed nanomaterials, being mixed with a binder, can be further used for production of highly adhesive ultrablack coatings for optical devices where even weak undesired reflections represent crucial issue.

\section*{Conflicts of interest}
There are no conflicts to declare.

\section*{Acknowledgements}
Laser-related experiments were supported by Russian Science Foundation (grant no. 19-79-00214). A.K. and S.M. express their gratitude to the Ministry of Science and Higher Education of the Russian Federation (grants nos. MÊ-3258.2019.8 and MK-3514.2019.2) regarding performed calculations of temperature profiles and light-to-heat conversion efficiency.

\end{document}